\def\beq{\begin{equation}}
\def\eeq{\end{equation}}
\def\beqn{\begin{eqnarray}}
\def\eeqn{\end{eqnarray}}

\def\vev#1{\left\langle #1\right\rangle}
\def\l{\langle}
\def\r{\rangle}

\def\l{\langle}
\def\r{\rangle}

\def\inbar{\,\vrule height1.5ex width.4pt depth0pt}

\def\IC{\relax\hbox{$\inbar\kern-.3em{\rm C}$}}
\def\IQ{\relax\hbox{$\inbar\kern-.3em{\rm Q}$}}
\def\IR{\relax{\rm I\kern-.18em R}}
 \font\cmss=cmss10 \font\cmsss=cmss10 at 7pt
\def\IZ{\relax\ifmmode\mathchoice
 {\hbox{\cmss Z\kern-.4em Z}}{\hbox{\cmss Z\kern-.4em Z}}
 {\lower.9pt\hbox{\cmsss Z\kern-.4em Z}}
 {\lower1.2pt\hbox{\cmsss Z\kern-.4em Z}}\else{\cmss Z\kern-.4em Z}\fi}

\def\AEF{Faraggi A E }
\def\NPB#1#2#3{ 19#2  {\it Nucl.\ Phys.\ B}\/ {\bf #1}  #3}
\def\PLB#1#2#3{ 19#2  {\it Phys.\ Lett.\ B}\/ {\bf #1}  #3}
\def\PRD#1#2#3{ 19#2  {\it Phys.\ Rev.\ D}\/ {\bf #1}  #3}

\def\IJMP#1#2#3{ 19#2  {\it Int.\ J.\ Mod.\ Phys.\ A}\/ {\bf #1} #3}

\font\bigbf=cmssbx10 scaled\magstep2

\font\it=cmti10 at 12pt
\font\ss=cmss10 at 12pt

\ss                                     
\documentstyle[iopconf1]{article}
\begin{document}
\rightline{UFIFT--HEP--97/21}
\rightline{hep-ph/9707311}
\rightline{July 1997}
\title{~~~~~~~~~~~~~~~~~Superstring Phenomenology\\
        ~~~~~~~~~~~~~~~~~~~~Present--and--Future\\
        ~~~~~~~~~~~~~~~~~~~~~~~~~Perspective\footnote{Invited talk presented 
at Beyond the Desert 97, Castle Ringberg, Tegernsee, Germany, 8-14 June 1997.}}

\author{Alon E. Faraggi\dag\footnote{E-mail:
faraggi@phys.ufl.edu.}}

\affil{\dag\ Department of Physics, University of Florida, 
Gainesville, FL 32611}

\beginabstract
The objective of superstring phenomenology is to develop
the models and methodology needed to connect quantitatively
between Planck scale physics and electroweak scale experimental
data. I review the present status of this endeavor
with a focus on the three generation free fermionic models.
\endabstract
\section{Introduction}

Superstring phenomenology aims at achieving two goals.
The first is to reproduce the physics of the Standard Model.
The second is to identify possible experimental signature of superstring
unification which may provide further evidence for its validity.
A model which
satisfies all of the experimental constraints,
is likely to be more than an accident. Such a model,
or class of models, will then serve as the laboratory for the
search for exotic predictions of superstring unification, and
as a laboratory in which we can address the
important question of how the string vacuum is selected.
I would like to remark that the goal of superstring
phenomenology should not necessarily be viewed as ``to derive
the correct string vacuum'', but to develop the models and
methodology needed to connect quantitatively
between physical phenomena, which
in the framework of unification is believed to originate at the
Planck scale, and experimental data at the electroweak scale. 
One should not forget that physics is primarily an experimental science, 
and as current data seem to support the existence of unification, 
this endeavor is required.

The first question that we must ask is why and whether the pursuit of
superstring theory is justified. 
After all
one should not overlook the fact that currently it is
assumed that gravity and the gauge interactions
become compatible only near the Planck scale, which is some
seventeen orders of magnitude above the scale were experiments
are performed. Thus, superstring unification, based on input from
current experiments, necessarily ignores the possibility that the
real physics above the electroweak scale is entirely different from
what we may have naively envisioned. 

Ever since the Pati--Salam \cite{patisalam} and the Georgi--Glashow
\cite{gg} seminal papers, we
know that the gauge groups and matter content of the Standard--Model
fit into representations of larger unifying 
gauge groups. This is an amazing coincidence that
correlates at least eighteen experimental observable, if one just counts 
the quantum numbers of the Standard Model matter states. 
The match is so striking that
it is hard to believe that nature is merely playing
a cruel mirage. Proton lifetime constraints impose that such higher
unification must occur at a scale which is at least of the order
$10^{16-17}{\rm GeV}$, just one or two orders of magnitude below the
Planck scale. Furthermore, low energy measurements of the Standard Model
gauge coupling are in qualitative agreement with the idea of
unification \cite{gqw,gcumssm}, 
while precision measurements of the Standard Model parameters
at the electroweak scale impose severe constraints on various extensions
of the Standard Model. Notable among this extensions is weak scale
supersymmetry, an important component in the program of unification, 
which continuous to be in agreement with the experimental data.
Thus, it is reasonable to hold the view that current low energy data
support the notion of the big desert scenario and unification. 

While unification and supersymmetry
are very attractive concepts they fall short of being satisfactory.
Many fundamental questions still cannot be addressed within their 
domain. The first type of questions is that supersymmetry and
unification cannot answer why the Standard Model has the 
structure that it has. Namely, what is the mechanism that selects
the gauge group and the matter content of the Standard Model.
Furthermore, although some of the low energy couplings are
manifestation of unified couplings at the high energy scale,
unification and supersymmetry, in general, cannot explain
the observed fermion mass spectrum. Most important,
traditional point quantum field theories and gravity are
incompatible. Thus, although gauge unification occurs
one or two orders of magnitude below the Planck scale, gravity
is still not incorporated. This is a crucial drawback of
conventional grand unified theories as gravitational effects
start to play a deterministic role near the unification scale. 

In the context of unified theories the solution to
these problems will come from some fundamental theory
at the Planck scale where gravity and quantum gauge
field theories become of comparable strength. This
elusive theory will hopefully tell us how the 
parameters of the Standard Model obtain their values. To date, 
superstring theory is the most developed Planck scale theory. 
It is believed to provide a framework for the consistent
unification of gravity and the gauge interactions. As such
it is currently the best probe that we have to explore physics
at the Planck scale. It therefore makes sense to try to see
if it is possible to connect superstring theory with the Standard
Model. This is the subject of superstring phenomenology. 
\section{Different approaches to superstring phenomenology}

There are currently two complementary approaches to superstring
phenomenology. In the first, which can be regarded as a view from above, 
the general strategy is to first try to understand what is the
nonperturbative formulation of string theory. The hope is that
the string vacuum will then be uniquely fixed and the low energy predictions
unambiguously determined. The second, which can be regarded as the 
view from below, asserts that we must use low energy data to single
out phenomenologically interesting superstring vacua. Such string 
models will then be instrumental to understand the dynamics
which select the string vacuum. These two approaches are in a sense
complementary and progress is likely to be made by pursuing
both approaches in parallel. Another view which I heard from
Henry Tye suggests that just as our earth and our sun
are not at the center of the universe, perhaps there does
not exist a unique string vacuum and in effect we happen to live 
in a typical phenomenological string vacuum. If this is the correct
view then while it still makes sense to look for the fundamental 
physical principles which underly string theory, or in more general
terms, which underly Planck scale physics, string dynamics may
not be able to reveal to us what is the superstring vacuum of our world.

I think that the discussion in the previous paragraph
demonstrates a different point in regard to the present status of
superstring phenomenology. 
The point is that the inclination toward one approach or another
arises from the age old question of ``what is the right question ?''
That is what is the question whose answer will lead
to real progress in understanding the objective structure of the
fundamental particles and forces. Some people believe that the right
question is ``What is string theory ?'', and hence what is the fundamental
non perturbative formulation of string theory. Others may believe that the
important question is ``how does unification occur ?'' and regard, for example,
the MSSM gauge coupling unification as strong support for addressing
this question. Yet another possibly important question is ``how
is supersymmetry broken ?''.  
Another point of view
is expressed concisely in a recent paper of Pati \cite{pati}.
In this view the important
questions are ``how can the proton be stable ?'', and ``how can the neutrino
masses be so small ?''. The point is that if we assume that unification
exist, then generically unification gives rise to many sources for
proton decay and may also generate large neutrino masses. Experimentally
the proton lifetime and the neutrino masses are suppressed to many
orders of magnitude. Thus, these two experimental observations
provide the most severe constraints on any model of unification.
To satisfy those constraints simultaneously in a satisfactorily robust
way are in this view the important questions to address. 

In non supersymmetric grand unification proton decay is induced 
by the gauge bosons of the grand unifying gauge group. The experimental
constraints then restrict that the unification scale, and the masses
of the grand unifying gauge bosons, have to be above roughly $10^{16}$GeV. 
The naive models then cannot be simultaneously consistent with
the gauge coupling unification and proton stability. Complicating the
spectrum may assist by pushing the unification to a higher scale.
This is precisely what supersymmetry does. Supersymmetry, however, 
introduces many new sources of proton decay through dimension four and five
operators. These new channels, in a generic supersymmetric model, will
cause the proton to decay extremely fast. Thus, very severe constraints
must be imposed on these new supersymmetric operators. In field theoretic
supersymmetric theories this is achieved by imposing $R$--parity, which
forbids dimension four operators, while a non trivial doublet--triplet
splitting mechanism is needed to evade proton decay from dimension five
operators. Thus, in field theoretic supersymmetric theories it may be
possible to circumvent the problem. However, the type of symmetries
that must be imposed are not expected to survive once we try to embed
supersymmetry into a theory that unifies gravity with the gauge interactions.
In gravity unified theories,
to paraphrase one of the speakers in this
conference, ``everything which is not
explicitly forbidden is allowed''.
Similarly the field theoretic doublet--triplet splitting mechanism will
be difficult, if not impossible, to implement in a theory that unifies 
gravity with the gauge interactions. 
Gravity unified theories would generically also give rise
to operators which induce large neutrino masses, far above 
the experimental limits \cite{fp}. 

A search for a suitable resolution of these issues
is likely to single out the phenomenologically interesting
models of unification. Superstring models offer appealing solutions to
these problems. In regard to proton stability superstring theory
allows the existence of unification without the need for an enlarged
grand unifying gauge group in the effective field theory level.
In this respect we may have a grand unifying gauge group, which
is broken at the string level
rather than in the effective field theory level. This is 
an appealing situation because in that case the doublet--triplet splitting
problem is resolved by a superstringy doublet--triplet splitting
mechanism \cite{ps}.
In this superstringy mechanism color triplets that mediate rapid 
proton decay are simply projected out from the massless spectrum by 
the generalized GSO projections. Thus, dimension five operators
can only be induced by heavier string modes. In specific models
it is also possible to show that also heavy string modes do
not induce such operators and in these models proton decay from
dimension five operators is entirely forbidden. As for
dimension four operators, again in the generic situation they
are abundant and are difficult to avoid. The point is that
higher order nonrenormalizable terms may induce dimension four
operators that are not sufficiently suppressed. Their absence can
only be insured by a gauge symmetry or a local discrete symmetry
that will guarantee that the nonrenormalizable terms do not
appear to a sufficient high order. 
\section{Superstring constructions}

There are several possible ways to try to construct
realistic superstring models. One possibility is to
construct superstring models with an intermediate 
GUT gauge group, like $SU(5)$, $SO(10)$, $E_6$, etc,
which is broken to the Standard Model
gauge group at an intermediate energy scale \cite{stringguts}.
This option imposes that the intermediate non--Abelian gauge group is
realized as a higher level affine Lie algebra because level one models
do not contain adjoint Higgs representations in the massless spectrum. 

The second possibility is to construct superstring models with semi--simple
GUTs, like $SU(3)^3$ \cite{suthree}, $SU(5)\times U(1)$
\cite{revamp,flipped}
or $SO(6)\times SO(4)$ \cite{patisalamstrings}.
In these models the extended non--Abelian symmetry
is broken in the effective field theory at an intermediate 
energy scale. This type of models can be realized with $k\ge1$
affine Lie algebra. 

The last possibility is to construct superstring models in which
the non--Abelian factors of the Standard--Model gauge group are obtained
directly at the string levels \cite{zthree,otherrsm,fny,eu,top,slm}.
These models again can be realized 
with $k\ge1$ affine Lie algebra. I propose that proton decay
constraints suggest that this last possibility is the correct choice. 

\smallskip
The general goal is therefore to construct superstring models
that are as realistic as possible. A realistic model of unification
must satisfy a large number of constraints
a few of which are listed below.

\centerline{{$\underline{{\hbox{~~~~~~~~~~~~~~~~~~~~~~~~~~~~~~~~~~~~~}}}$}}

{}~~~~~$1.$ Gauge group ~$\longrightarrow$~ $SU(3)\times SU(2)\times U(1)_Y$

{}~~~~~$2.$ Contains three generations

{}~~~~~$3.$ Proton stable ~~~~~~~~($\tau_{\rm P}>10^{30+}$ years)

{}~~~~~$4.$ N=1 supersymmetry~~~~~~~~(or N=0)

{}~~~~~$5.$ Contains Higgs doublets $\oplus$ potentially realistic
Yukawa couplings

{}~~~~~$6.$ Agreement with $\underline{\sin^2\theta_W}$ and
$\underline{\alpha_s}$ at $M_Z$ (+ other observables).

{}~~~~~$7.$ Light left--handed neutrinos

{}~~~~~~~~~~~~~$~8.$ $SU(2)\times U(1)$ breaking

{}~~~~~~~~~~~~~$~9.$ SUSY breaking

{}~~~~~~~~~~~~~$10.$ No flavor changing neutral currents

{}~~~~~~~~~~~~~$11.$ No strong CP violation

{}~~~~~~~~~~~~~$12.$ Exist family mixing and weak CP violation

{}~~~~~$13.$ +~~ {\bf ...}

{}~~~~~$14.$ +~~~~~~~~~~~~~~~~{\bigbf{GRAVITY}}

\centerline{{$\underline{{\hbox{~~~~~~~~~~~~~~~~~~~~~~~~~~~~~~~~~~~~~}}}$}}

\smallskip

The question then is whether it is possible
to construct a model which satisfies all of those criteria, or
possibly a class of models which can accommodate most of these
constraints. To date the most developed theory that can
consistently unify gravity with the gauge interactions is string
theory. While alternatives may exist, it makes sense
at this stage to try to use string theory to construct a model which
satisfies the above requirements. Even if eventually string theory
turns out not to be the fundamental theory of nature, a model which
satisfies all of above constraints is likely to arise as an effective
model from the true fundamental theory.

There are five known superstring theories in ten dimensions. The type I,
type II A\&B and the $E_8\times E_8$ and $SO(32)$ heterotic strings. 
All these 10 dimensional string theories and including the 11 dimensional 
supergravity are believed to be special limits of one underlying theory. 
This is the understanding which emerges from the various string dualities
that were uncovered in the last couple of years. The formulation of the
underlying theory is still unclear and the hope is that it will improve 
our understanding of the mechanism which selects the string vacuum 
in four dimensions. In terms of trying to connect superstring theory
with experimental physics the best that we can do at the moment is
to continue to study promising four dimensional vacua and keep an
open eye on progress in the formal understanding of superstring theory. 

The exploration of realistic superstring vacua proceeds by studying
compactification of the heterotic string from ten to four dimensions. 
There is a large number of possibilities. The first type of semi--realistic
superstring vacua that were constructed are compactification
on Calabi--Yau manifolds which give rise to an $E_6$ observable
gauge group which is broken further by the Hosotani flux breaking
mechanism to $SU(3)^3$ \cite{suthree}.
This gauge group is then broken to the 
Standard Model gauge group in the effective field theory level. 
This type of geometrical compactifications are not exact conformal
field theories and correspond at special points to
conformal theories which have $N=2$ world--sheet supersymmetry
in the left and right moving sectors. Similar geometrical compactifications
which have only (2,0) world--sheet supersymmetry have also been
studied and can lead to compactifications with $SO(10)$ and $SU(5)$
observable gauge group \cite{twozero}.
The analysis of this type of compactification
is complicated due to the fact that they do not correspond
to free world--sheet theories. Therefore, it is rather difficult
to try to calculate the parameters of the Standard Model
in these compactifications. On the other hand they can provide
a more sophisticated mathematical window to the underlying quantum geometry
and indeed much of the effort in this direction has focused on the formal
development of the underlying quantum geometry \cite{greene}. 

The next class of superstring vacua that have been explored in detail are
the orbifold models \cite{dhvw}.
In these models one starts with a compactification
of the heterotic string on a flat torus,
using the Narain prescription \cite{narain}.
This type of compactification uses free world--sheet bosons. The Narain
lattice is then moded out by some discrete symmetries which are the
orbifold twisting. The most detailed study of this type of models are
the $Z_3$ orbifold \cite{zthree}
which give rise to three generation models with
$SU(3)\times SU(2)\times U(1)^n$ gauge group. One caveat of this class of
models is that the weak--hypercharge does not have the standard
$SO(10)$ embedding.
Thus, the nice features of $SO(10)$ unification are lost. This fact has
a crucial implication that the normalization of the weak hypercharge
relative to the non--Abelian currents is larger than 5/3, the standard
$SO(10)$ normalization. This results generically in disagreement with the
observed low energy values for $\sin^2\theta_W(M_Z)$ and $\alpha_s(M_Z)$. 

Another special type of string compactifications that has been
studied in detail are the free fermionic models. The simplest
examples correspond to $Z_2\times Z_2$ orbifolds at special points
in the compactification space. These models give rise to the most
realistic superstring models constructed to date. There are several
key features of these models that suggest that if string
theory is relevant in nature, then the true string vacuum is 
in the vicinity of these models. First is the fact that the free
fermionic models are formulated at a highly symmetric point in the
string compactification space. The second is that in the $Z_2\times Z_2$
orbifold models at the free fermionic point the emergence of three
generations is correlated with the underlying structure of the orbifold
model \cite{ztwo}. Each of the Standard Model generations is obtained from
one of the twisted sectors and carries horizontal charges under
one of the orthogonal planes of the $Z_2\times Z_2$ orbifold model.
Naively one may view the existence of three generations in nature
as arising because we are dividing the six dimensional compactified
space into factors of two.  
Thus, these models may explain the existence of three generation
in nature in terms of the underlying geometry. The last important point
is that in the free fermionic models the weak hypercharge has
the standard $SO(10)$ embedding. Consequently these models can be
in agreement with the observed low energy values for 
$\sin^2\theta_W(M)$ and $\alpha_s(M_Z)$. This class of models
and their phenomenology are the focus of this talk. 
\section{Free fermionic models}

In the free fermionic formulation \cite{FFF}
for the left--movers  (world--sheet supersymmetric) one has the 
usual space--time fields $X^\mu$, $\psi^\mu$, ($\mu=0,1,2,3$), 
and in addition the following eighteen real free fermion fields:
$\chi^I,y^I,\omega^I$  $(I=1,\cdots,6)$, transforming as the adjoint 
representation of $SU(2)^6$. The supercurrent 
is given in terms of these fields as follows
$$T_F(z) = \psi^\mu\partial_zX_\mu + {\sum_{i=1}^6}\chi^iy^i\omega^i.$$
For the right movers we have ${\bar X}^\mu$ and 44 real free fermion fields:
${\bar\phi}^a$, $a=1,\cdots,44.$
A model in this construction 
is defined by a set of  basis vectors of boundary conditions for all 
world--sheet fermions, which are constrained by the string 
consistency requirements and 
completely determine the vacuum structure of
the model. The physical spectrum is obtained by applying the generalized 
GSO projections. The low energy effective field theory is obtained
by S--matrix elements between external states. The Yukawa couplings 
and higher order nonrenormalizable terms in the superpotential
are obtained by calculating correlators between vertex operators.
For a correlator to be non vanishing all the symmetries of the model must 
be conserved. Thus, the boundary condition vectors determine the  
phenomenology of the models.

The first five vectors in the basis of the models that I discuss here
consist of the NAHE{\footnote{This set was first 
constructed by Nanopoulos, Antoniadis, Hagelin and Ellis  (NAHE) 
in the construction
of  the flipped $SU(5)$.  {\it nahe}=pretty, in
Hebrew.}} set. 
The gauge group after the NAHE set is $SO(10)\times
E_8\times SO(6)^3$ with $N=1$ space--time supersymmetry, 
and 48 spinorial $16$ of $SO(10)$, sixteen from each sector 
$b_1$, $b_2$ and $b_3$. The NAHE set divides the internal world--sheet 
fermions in the following way: ${\bar\phi}^{1,\cdots,8}$ generate the 
hidden $E_8$ gauge group, ${\bar\psi}^{1,\cdots,5}$ generate the $SO(10)$ 
gauge group, and $\{{\bar y}^{3,\cdots,6},{\bar\eta}^1\}$,  
$\{{\bar y}^1,{\bar y}^2,{\bar\omega}^5,{\bar\omega}^6,{\bar\eta}^2\}$,
$\{{\bar\omega}^{1,\cdots,4},{\bar\eta}^3\}$ generate the three horizontal 
$SO(6)^3$ symmetries. The left--moving $\{y,\omega\}$ states are divided 
to $\{{y}^{3,\cdots,6}\}$,  
$\{{y}^1,{y}^2,{\omega}^5,{\omega}^6\}$,
$\{{\omega}^{1,\cdots,4}\}$ and $\chi^{12}$, $\chi^{34}$, $\chi^{56}$ 
generate the left--moving $N=2$ world--sheet supersymmetry. 

The internal fermionic states $\{y,\omega\vert{\bar y},{\bar\omega}\}$
correspond to the six left--moving and six right--moving compactified 
dimensions in a geometric formulation. This correspondence is illustrated
by adding the vector
$X$ to the NAHE set, with periodic boundary conditions for the set 
$({{\bar\psi}^{1,\cdots,5}},{{\bar\eta}^{1,2,3}})$ and antiperiodic
boundary conditions for all other world--sheet fermions. 
This boundary condition vector extends the gauge symmetry to 
$E_6\times U(1)^2\times E_8\times SO(4)^3$ with $N=1$ supersymmetry 
and twenty-four chiral $27$ of $E_6$. The same model is generated in the 
orbifold language \cite{dhvw} 
by moding out an $SO(12)$ lattice by a $Z_2\times{Z_2}$ 
discrete symmetry with standard embedding. In the construction of 
the standard--like models beyond the NAHE set, the assignment 
of boundary conditions to the set of internal fermions 
$\{y,\omega\vert{\bar y},{\bar\omega}\}$ determines many of the 
properties of the low energy spectrum, such as the number of 
generations, the presence of Higgs doublets, Yukawa couplings, etc. 

In the realistic free fermionic models the boundary condition vector $X$
is replaced by the vector $2\gamma$ in which $\{{\bar\psi}^{1,\cdots,5},
{\bar\eta}^1,{\bar\eta}^2,{\bar\eta}^3,{\bar\phi}^{1,\cdots,4}\}$
are periodic and the remaining left-- and right--moving fermionic states 
are antiperiodic. The set $\{1,S,2\gamma,\xi_2\}$ generates a model 
with $N=4$ space--time supersymmetry and $SO(12)\times SO(16)\times SO(16)$
gauge group. The $b_1$ and $b_2$ twists are applied to reduce
$N=4$ to $N=1$ space--time supersymmetry. 
The gauge group is broken to  $SO(4)^3\times U(1)^3\times SO(10)\times E_8$.
The $U(1)$ combination $U(1)=U(1)_1+U(1)_2+U(1)_3$ has a non--vanishing
trace and the trace of the two orthogonal combinations vanishes.
The number of generations is still 24, eight from each sector $b_1$, $b_2$
and $b_3$. The chiral generations are now $16$ of $SO(10)$ from the sectors
$b_j$ $(j=1,2,3)$. The $10+1$ and the $E_6$
singlets from the sectors $b_j+X$ are replaced by vectorial
$16$ of the hidden $SO(16)$ gauge group from the sectors
$b_j+2\gamma$. As I will show below the structure of the 
sector $b_j+2\gamma$ with respect to the sectors $b_j$ plays an important
role in the texture of fermion mass matrices. 
The anomalous $U(1)$ in this models is seen to arise due to
the breaking of $(2,2)\rightarrow(2,0)$ world--sheet supersymmetry.
This anomalous $U(1)$ generates a Fayet--Iliopoulos D--term \cite{dsw}
which breaks supersymmetry at the Planck scale. Supersymmetry
is restored by assigning non vanishing VEVs to a set of Standard
Model singlets in the massless string spectrum along flat F and D
directions. 

The standard--like models are constructed by adding three additional 
vectors to the NAHE set \cite{fny,eu,top,slm}. The $SO(10)$ symmetry
is broken in two stages, first to $SO(6)\times SO(4)$ and next to
$SU(3)\times SU(2)\times U(1)^2$. At the same time the number of
generations is reduced to three generations one from each sector
$b_1$, $b_2$ and $b_3$. The flavor $SO(6)^3$ symmetries are broken
to factors of $U(1)_{R_j}$ and $U(1)_{R_{j+3}}$ $(j=1,2,3)$. For every
right--moving gauged $U(1)$ symmetry there is a corresponding left--moving
global $U(1)$ symmetry $U(1)_{L_j}$ and $U(1)_{L_{j+3}}$. Finally, each
generation is charged with respect to two Ising model operators which are
obtained by pairing a left--moving real fermion with a right--moving real
fermion. The Higgs spectrum consists of three pairs from the Neveu--Schwarz
sector, ${h_j}$ and ${\bar h}_j$, with charges under the
horizontal symmetries and one or two additional pairs from the sector
$b_1+b_2+\alpha+\beta$. Analysis of the Higgs mass matrix suggests that
at low energies only one pair of Higgs multiplets remains light.
The pair which remains light, together with the flavor symmetries
which are broken near the string scale, then determines the hierarchical
structure of the fermion mass matrices.
\section{Fermion masses}

One of the important achievements of the
free fermionic models is the
successful prediction of the top quark mass and
the explanation of the heaviness of the top quark mass
relative to the masses of the lighter quarks and leptons.
The cubic level Yukawa couplings for the quarks and leptons are 
determined by the boundary conditions in the vector $\gamma$
according to the following rule \cite{slm}
\begin{eqnarray}
\Delta_j&&=
\vert\gamma(U(1)_{L_{j+3}})-\gamma(U(1)_{R_{j+3}})\vert=0,1
{\hskip 1cm}(j=1,2,3)\\
\Delta_j&&=0\rightarrow d_jQ_jh_j+e_jL_jh_j; \\
\Delta_j&&=1\rightarrow u_jQ_j{\bar h}_j+N_jL_j{\bar h}_j,
\end{eqnarray}
 where $\gamma(U(1)_{R_{j+3}})$, $\gamma(U(1)_{L_{j+3}})$ are the boundary 
conditions of the world--sheet fermionic currents that generate the  
$U(1)_{R_{j+3}}$, $U(1)_{L_{j+3}}$ symmetries. 
In models with $\Delta_{1,2,3}=1$ only $+{2\over3}$ charged
type quarks get a cubic level Yukawa coupling. 

Such models therefore suggest an explanation for the top quark mass hierarchy 
relative to the lighter quarks and leptons. At the cubic level only the 
top quark gets a mass term and the mass terms for the lighter 
quarks and leptons are obtained from nonrenormalizable terms. 
To study this scenario we have to examine the nonrenormalizable 
contributions to the doublet Higgs mass matrix and to the fermion mass
matrices \cite{NRT,CKM}. 

At the cubic level there are two pairs of electroweak doublets. 
At the nonrenormalizable level one additional pair receives a superheavy 
mass and one pair remains light to give masses to the fermions at 
the electroweak scale. Requiring F--flatness imposes that the light 
Higgs representations are ${\bar h}_1$ or ${\bar h}_2$ and $h_{45}$. 

The nonrenormalizable fermion mass terms of order $N$ are of the form 
$cgf_if_jh\phi^{^{N-3}}$ or
$cgf_if_j{\bar h}\phi^{^{N-3}}$, where $c$ is a 
calculable coefficient, $g$ is the gauge coupling at the unification 
scale,  $f_i$, $f_j$ are the fermions from
the sectors $b_1$, $b_2$ and $b_3$, $h$ and ${\bar h}$ are the light 
Higgs doublets, and $\phi^{N-3}$ is a string of Standard Model singlets
that get a VEV and produce a suppression factor
${({{\langle\phi\rangle}/{M}})^{^{N-3}}}$ relative to the cubic
level terms. Several scales contribute to the generalized VEVs. The 
leading one is the scale of VEVs that are used to cancel 
the ``anomalous'' $U(1)$
D--term equation. The next scale is generated by Hidden sector 
condensates. Finally, there is a scale which is related to the breaking
of $U(1)_{Z^\prime}$ which is embedded in $SO(10)$ and
is orthogonal to the weak hypercharge.

At the cubic level only the top quark gets a non vanishing mass term. 
Therefore only the top quark mass is characterized by the electroweak 
scale. The remaining quarks and leptons obtain their mass terms from 
nonrenormalizable terms. The cubic and nonrenormalizable terms in the 
superpotential are obtained by calculating correlators between the vertex 
operators. The top quark Yukawa coupling is generically given by,
$g\sqrt2$,
where $g$ is the gauge coupling at the unification scale. In 
the model of Ref. \cite{top}, bottom quark and tau lepton mass terms 
are obtained at the quartic order,
$$W_4=\{{d_{L_1}^c}Q_1h_{45}^\prime\Phi_1+{e_{L_1}^c}L_1h_{45}^\prime\Phi_1+
{d_{L_2}^c}Q_2h_{45}^\prime{\bar\Phi}_2
+{e_{L_2}^c}L_2h_{45}^\prime{\bar\Phi}_2\}.$$
The VEVs of $\Phi$ are obtained from the cancelation of the anomalous 
$U(1)$ D--term equation. The coefficient of the quartic order mass terms 
were calculated by calculating the quartic order correlators and the 
one dimensional integral was evaluated numerically. Thus after 
inserting the VEV of ${\bar\Phi}_2$ the effective bottom quark and tau lepton
Yukawa couplings are given by \cite{top}, 
$$\lambda_b=\lambda_\tau=0.35g^3.$$
They are suppressed relative to the top Yukawa by
$${{\lambda_b}\over{\lambda_t}}=
{{0.35g^3}\over{g\sqrt2}}\sim{1\over8}.$$
To evaluate the top quark mass, the three Yukawa couplings are run 
to the low energy scale by using the MSSM RGEs. The bottom mass is 
then used to calculate $\tan\beta$ and the top quark mass is 
found to be \cite{top},
$$m_t\sim175-180GeV.$$
The fact that the top Yukawa is found near a fixed point suggests that this
is in fact a good prediction of the superstring standard--like models.
By varying $\lambda_t\sim0.5-1.5$ at the unification
scale, it is found that $\lambda_t$ is always $O(1)$ at the 
electroweak scale. 

The analysis of the fermion masses of the lighter quarks and
leptons proceeds by analyzing higher order nonrenormalizable terms.
An analysis of fermion mass terms up to order $N=8$ revealed the general 
texture of fermion mass matrices in these models \cite{NRT}.
The sectors $b_1$ 
and $b_2$ produce the two heavy generations.
Their mass terms are suppressed by singlet VEVs that are used in the 
cancelation of the anomalous $U(1)$ D--term equation. 
The sector $b_3$ produces the lightest generation. 
The diagonal 
mass terms for the states from $b_3$ can only be generated by 
VEVs that break $U(1)_{Z^\prime}$. This is due to the horizontal 
$U(1)$ charges and because the Higgs pair $h_3$ and ${\bar h}_3$ 
necessarily gets a Planck scale mass \cite{NRT}. 
The suppression of the lightest generation mass terms is seen to be 
a result of the structure of the vectors $\alpha$ and $\beta$
with respect to the sectors $b_1$, $b_2$ and $b_3$. 
The mixing between the generations
is obtained from exchange of states from the sectors $b_j+2\gamma$. The 
general texture of the fermion mass matrices in the superstring 
standard--like models is of the following form, 
$${M_U\sim\left(\matrix{\epsilon,a,b\cr
                    {\tilde a},A,c \cr
                    {\tilde b},{\tilde c},\lambda_t\cr}\right);{\hskip .2cm}
M_D\sim\left(\matrix{\epsilon,d,e\cr
                    {\tilde d},B,f \cr
                    {\tilde e},{\tilde f},C\cr}\right);{\hskip .2cm}
M_E\sim\left(\matrix{\epsilon,g,h\cr
                    {\tilde g},D,i \cr
                    {\tilde h},{\tilde i},E\cr}\right)},$$
where $\epsilon\sim({{\Lambda_{Z^\prime}}/{M}})^2$.
The diagonal terms in capital letters represent leading 
terms that are suppressed by singlet VEVs, and $\lambda_t=O(1)$.
The mixing terms are generated by hidden sector states from the
sectors $b_j+2\gamma$ and are represented by small letters. They 
are proportional to $({{\langle{TT}\rangle}/{M}^2})$.   

In Ref. \cite{CKM} it was shown that if the states from the
sectors $b_j+2\gamma$ obtain 
VEVs in the application of the DSW mechanism, then a Cabibbo angle of the 
correct order of magnitude can be obtained in the superstring standard--like 
models. For one specific choice of singlet VEVs that solves the cubic
level F and D constraints the down mass matrix $M_D$ is given by
\begin{equation}
M_d\sim\left(\matrix
{&\epsilon
&{{V_2{\bar V}_3\Phi_{45}}\over{M^3}} &0\cr
&{{V_2{\bar V}_3\Phi_{45}\xi_1}\over{M^4}} 
&{{{\bar\Phi}_2^-\xi_1}\over{M^2}} &0 \cr
&0 &0 
&{{\Phi_1^+\xi_2}\over{M^2}}\cr}\right)v_2,
\label{cabmix}
\end{equation}
where $v_2=\l h_{45} \r$ and we have used  
${1\over2}g\sqrt{2\alpha^\prime}=\sqrt{8\pi}/M_{Pl}$, to define
$M\equiv M_{Pl}/2\sqrt{8\pi}\approx 1.2\times 10^{18}GeV$ \cite{KLN}.
The undetermined VEVs of $\bar \Phi_{13}$ and $\xi_2$ are used to fix $m_b$ and
$m_s$ such that $\l \xi_1 \r \sim M$. We also take $tan \beta=v_1/v_2 
\sim 1.5$.
Substituting the values of the VEVs above and 
diagonalizing $M_D$ by a bi--unitary transformation we obtain the 
Cabibbo mixing matrix
\begin{equation}
\vert V \vert\sim \left(\matrix {0.98&0.2&0 \cr 
                                        0.2&0.98&0 \cr 
                                        0&0&1 \cr } \right). 
\label{cabmixmat}
\end{equation}
Since the running from the scale $M$ down to the weak scale does not affect
the Cabibbo angle by much \cite{GLT}, we conclude that realistic mixing of the 
correct order of magnitude can be obtained in this scenario.
The analysis was extended to show that reasonable 
values for the entire CKM matrix parameters can be obtained for appropriate
flat F and D solutions. For one specific solution the up and down quark mass 
matrices take the form 
\begin{equation}
M_u\sim\left(\matrix{&\epsilon
&{{V_3{\bar V}_2\Phi_{45}\bar \Phi_3^+}\over{M^4}} &0\cr
&{{V_3{\bar V}_2\Phi_{45}\bar \Phi_2^+}\over{M^4}} 
&{{{\bar\Phi}_i^-\bar \Phi_i^+}\over{M^2}} 
&{V_1{\bar V}_2\Phi_{45}\bar \Phi_2^+}\over{M^4} \cr
&0 &{V_1{\bar V}_2\Phi_{45}{\bar\Phi}_1^+}\over{M^4} 
&1\cr}\right)v_1,
\label{mu}
\end{equation}
and
\begin{equation}
M_d\sim\left(\matrix{&\epsilon
&{{V_3{\bar V}_2\Phi_{45}}\over{M^3}} &0\cr
&{{V_3{\bar V}_2\Phi_{45}\xi_1}\over{M^4}} 
&{{{\bar\Phi}_2^-\xi_1}\over{M^2}} &{V_1{\bar V}_2\Phi_{45}\xi_i}\over{M^4} \cr
&0 &{V_1{\bar V}_2\Phi_{45}\xi_i}\over{M^4} 
&{{\Phi_1^+\xi_2}\over{M^2}}\cr}\right)v_2,
\label{md}
\end{equation}
with $v_1$, $v_2$ and $M$ as before.
Substituting the 
VEVs and diagonalizing $M_u$ and $M_d$ by a bi--unitary transformation, we 
obtain a qualitatively realistic mixing matrix.
The texture and hierarchy of the mass terms in Eqs.
(\ref{mu},\ref{md}) arise due to the set of singlet VEVs in the
solution to the $F$ and $D$--flatness constraints. 
The zeroes in the 13 and 31 entries of the mass matrices are protected to all 
orders of nonrenormalizable terms. To obtain a non--vanishing 
contribution to these entries either $V_1$ and ${\bar V}_3$ 
or $V_3$ and ${\bar V}_1$ must obtain a VEV simultaneously. Thus, 
there is a residual horizontal symmetry that protects these vanishing 
terms. The 11 entry in the mass matrices, e.g. the diagonal mass terms 
for the lightest generation states, can only be obtained from VEVs
that break $U(1)_{Z^\prime}$ \cite{NRT}. We assume that $U(1)_{Z^\prime}$ 
is broken at an intermediate energy scale that is suppressed relative 
to the scale of scalar VEVs \cite{NRT}. In Ref. \cite{FHn} we showed that 
$U(1)_{Z^\prime}$ is broken by hidden sector matter condensates 
at $\Lambda_{Z^\prime}\leq10^{14}GeV$. Consequently, 
we have taken $\epsilon\leq(\Lambda_{Z^\prime}/M)^2\sim10^{-8}$.  

Texture zeroes in the fermion mass matrices are 
obtained if the VEVs of some states from the sectors $b_j+2\gamma$ vanish. 
These texture zeroes are protected by the symmetries of the string models
to all order of nonrenormalizable terms \cite{CKM}.
For example in the above mass 
matrices the 13 and 31 vanish because $\{V_1,V_3\}$ get a VEV but ${\bar V}_1$
and ${\bar V}_3$ do not.
Other textures are possible for other choices of VEVs
for the states from the sectors $b_j+2\gamma$.  
\section{Gauge coupling unification}

An important issue in superstring  phenomenology is that of gauge
coupling unification. Superstring theory
predicts that the gravitational and gauge
couplings are unified and satisfy the relation,
\begin{equation}
{{8\pi G_N}\over {\alpha^\prime}}=g_i^2k_i\equiv g^2_{\rm string}
\end{equation}
where $G_N$ is the gravitational coupling, $\alpha^\prime$ is the Regge slope
and $k_i $ are the Kac-Moody level of the group factor $G_i $. 
Thus, superstring theory predicts that all gauge couplings are unified without
the need for a grand unified  gauge group. This is an important property 
of superstring theory. For while superstring theory still 
predicts unification of
the couplings, and for this reason many of the successful relations of grand 
unified theories can be retained, many of the problems associated with GUTs,
like too rapid proton decay etc.,
can be resolved. The string unification scale \cite{scales}
is of the order, 
\begin{equation}
M_{\rm string}\approx 5\times g_{string}\times 10^{17} {\rm GeV}
\end{equation}
If we naively assume that the spectrum below the string unification scale 
consists solely of the MSSM spectrum, {\it i.e.} three generations 
plus two Higgs doublets and the weak hypercharge normalization is 
$k_Y=5/3$, then the predicted values for $\sin^2\theta_W(M_Z)$ and
$\alpha_s(M_Z)$ are in contradiction with the experimentally observed
values. Alternatively we may identify the MSSM unification scale of
the order $2\times 10^{16} {\rm GeV}$ which is roughly a factor
of 20 below the string unification scale. This discrepancy is
known as the string gauge coupling unification problem. 

It would seem that in an extrapolation of the gauge couplings of 
fifteen orders of magnitude, there will be many possible resolutions
to this problem. In this regard we
must caution that one should view such fine structure of the theory
with a grain of salt. While we believe that our method of
extrapolation is consistent and safe, one should not forget that many
assumptions are being made with regard to the nature of the physics in the
desert between the weak and unification scales. With this remark
in mind, we find that in fact the string coupling unification problem is not 
easily resolved. To understand why this is the case, it suffices to examine 
the extrapolation of $\alpha_s$. We know that starting from the MSSM
unification scale and extrapolating to the $Z$ scale we get a value of order
$\alpha_s\approx0.12$, in agreement with the data. Now suppose that
we extrapolate further from the weak scale to the bottom scale.
The strong coupling at the bottom 
scale is then of order $0.22$, which agrees with the data. Now the effect of
starting the running at the string scale is exactly the same. Namely, 
instead of adding the additive interval between the $Z$ and the bottom scales,
we add it at the high scale between the MSSM and the string 
unification scales. Therefore, we find that $\alpha_s(M_Z)\approx 0.22$, which 
is roughly 100\% off the experimentally observed value. We see that 
although the discrepancy between the string and MSSM
unification scales is small, 
it translates into a large deviation in $\alpha_s$. This is the reason why
most of the sources that may affect the evolution of
$\alpha_s$ cannot, in fact, 
resolve the problem. 

In ref. \cite{df} the string gauge coupling unification problem was
examined in detail. From the gauge couplings renormalization group equations
we obtain solutions for $\sin^2\theta_W(M_Z)$ and $\alpha_s(M_Z)$, 
which include all the possible sources that may affect their values at
$M_Z$. These include: modification of the weak hypercharge normalization
\cite{hynormalization}, 
extended non-Abelian gauge group at an intermediate energy scale, 
threshold corrections due to the infinite tower of the heavy string modes, 
threshold corrections due to the light SUSY spectrum, the effect of 
additional matter thresholds at intermediate energy scales.
The analysis also includes the
two-loop gauge and Yukawa coupling
corrections to the central values, as well as
the effect of converting from the ${\overline{MS}}$  renormalization scheme
in which the parameters are measured to the ${\overline{DR}}$ scheme
in which the SUSY parameters are run. 

The most complex part of the analysis is the calculation of
the string threshold
corrections. This analysis is made complicated by the
fact that a typical realistic
superstring model contains a few hundred-thousand sectors over which we 
have to integrate. In addition one needs to worry about careful removal of the 
infrared divergences and the integration over the modular domain. Although
in simple unrealistic models it is possible to parameterize
the string threshold
corrections in terms of moduli fields
\cite{moduli}, this is in general not possible in the
realistic models which are by far more complicated.
A priori we have no reason
to expect that the string threshold corrections would be small.
It is a viable possibility that the string
thresholds would produce
large corrections that may shift the string unification
scale to the MSSM unification
scale. The calculation of the string threshold corrections is similar
to the calculation of the one--loop partition function. While in
the partition function we
count all states at each mass level, the string threshold corrections count
all states at each mass level times their lattice charge. 
Technical details of the calculation are given in ref. \cite{df}.

The conclusion of ref. \cite{df} is that of the 
perturbative corrections
to the one-loop RGEs, only the existence of additional
matter thresholds \cite{gaillard}
at intermediate energy scale may result in agreement of the string unification
scale with the low energy data. All other effects are either
too small or in fact
increase the value of $\alpha_s(M_Z)$ and therefore worsen the disagreement
with the low energy data. The effect of intermediate
color triplets is to sufficiently slow
down the evolution of $\alpha_s$, while all other effects are either too small
or increase $\alpha_s(M_Z)$. This is a remarkable conclusion as it shows
that perturbative string scale unification requires the
existence of additional matter
beyond the MSSM spectrum. String models generically give rise to additional 
matter beyond the MSSM spectrum, and some models in fact produce
the additional representations with the required weak hypercharge assignment, 
needed for the consistency of
string scale gauge coupling unification with the low 
energy data \cite{gcu}.
Another proposal \cite{witten} suggests that
nonperturbative effects in the context of M--theory are responsible
for shifting the string unification scale down to the MSSM unification 
scale. 
\section{Exotic matter}

The existence of additional matter beyond the MSSM spectrum may have 
important phenomenological and cosmological implications
\cite{ccf}. The extra
matter is often obtained in the realistic free fermionic models from 
sectors which arise due to the breaking of the $SO(10)$ symmetry to 
one of its subgroups. This matter can be classified according to 
the type of $SO(10)$ breaking in each sector. Sectors which break 
the $SO(10)$ symmetry to $SO(6)\times SO(4)$ or to $SU(5)\times U(1)$
give rise to states with fractional charge $\pm1/2$ while sectors which break
the $SO(10)$ symmetry directly to $SU(3)\times SU(2)\times U(1)^2$ 
give rise to states with the Standard Model charges but with fractional 
charges under $U(1)_{Z^\prime}$ which is embedded in $SO(10)$. 
Thus, for example we have states from this type of sectors which are
exotic leptoquarks \cite{elwood} 
{\it i.e.}, they have baryon number $\pm1/3$ but exotic
lepton number $\pm1/2$. Due to its exotic $U(1)_{Z^\prime}$ charge, 
in some models, it can be shown that the superpotential terms of such 
a state with the Standard Model states vanish to all
orders of nonrenormalizable terms. 
In this case the exotic states can interact with the Standard Model states
only via the gauge interactions and cannot decay into them . 
In such a model therefore an exotic state will be stable and 
one has to check that its mass density does not over close the universe. 
These constraints were investigated in detail in ref. \cite{ccf}.
Several general remarks however are in order. Exotic states
that do not have GUT origin are generic in superstring models. 
They arise due to the breaking of the non-Abelian gauge
symmetries at the string rather than in the effective field theory level.  
Such states are therefore a generic signature of superstring compactification. 
Thus, they may lead to possible observable experimental signatures.
For example, all the level one models predict the existence of fractionally 
charged states at least with Planck scale masses. Specific free fermionic
models also predict the existence of Standard Standard Model states which are 
exotic from the point of view of the underlying $SO(10)$. These may be 
color triplets, electroweak doublets or Standard Model singlets. Such states
may provide an experimental signature of specific classes of superstring
compactifications. 
\section{Supersymmetry breaking}

Next let me turn to the question of supersymmetry breaking. All the 
models that I discussed so far have $N=1$ supersymmetry at the Planck
scale. Supersymmetry must be broken at some scale as it is not observed
at the low scale.

we address the following question: Given a supersymmetric string 
vacuum at the Planck scale, is it possible to obtain hierarchical 
supersymmetry breaking in the observable sector? A supersymmetric 
string vacuum is obtained by finding solutions to the cubic level 
F and D constraints. We take a gauge coupling in agreement with 
gauge coupling unification, thus taking a fixed value for the dilaton VEV. 
We then investigate the role of nonrenormalizable terms and strong 
hidden sector dynamics. The hidden sector contains two non--Abelian 
hidden gauge groups, $SU(5)\times SU(3)$, with matter in vector--like 
representations. The hidden $SU(3)$ group is broken near the Planck scale. 
We analyze the dynamics of the hidden $SU(5)$ group. 
The  $SU(5)$ hidden matter mass matrix is given by
\begin{equation}
{\cal M}=\left(\matrix{ 0   & C_1   & 0   \cr 
                          B_1 & A_2   & C_2 \cr  
                          0   & C_3   & A_1   \cr  }\right)~,
\label{mbj2gamma}
\end{equation}
where $A,B,C$ arise from nonrenormalizable terms of orders $N=5,8,7$ 
respectively. A specific solution was found in ref. \cite{SUSYX} and
taking generically $\l\phi\r\sim{gM}/4\pi\sim M/10$ yielded $A_i \sim 
10^{15}~GeV$, $B_i \sim 10^{12}~GeV$, and $C_i\sim 10^{13}~GeV$. From 
Eqs. (\ref{mbj2gamma}) we observe that to insure a nonsingular
hidden matter mass matrix, we must require $C_1\ne0$ and $B_1\ne0$.
This imposes ${\bar V_3}\ne0$ and $V_2\ne0$. Thus, the nonvanishing VEVs 
that generate the Cabibbo mixing also guarantee the stability of the 
supersymmetric vacuum. 
The gaugino and matter
condensates are given by the well known expressions for supersymmetric $SU(N)$
with matter in $N+{\bar N}$ representations \cite{SQCD}, 
\begin{eqnarray}
{1\over{32\pi^2}}\vev{\lambda\lambda}
&&=\Lambda^3\left(det{{\cal M}\over\Lambda}\right)^{1/N},\nonumber\\
\Pi_{ij}=\vev{{\bar T_i}T_j}&&={1\over{32\pi^2}}\vev
{\lambda\lambda}{{\cal M}_{ij}}^{-1}\label{gmcondt}
\end{eqnarray}
where $\l\lambda\lambda\r$, $\cal M$ and 
$\Lambda$ are the hidden gaugino condensate, the hidden matter mass matrix 
and the $SU(5)$ condensation scale, respectively. 
Modular invariant generalization of Eqs. (\ref{gmcondt})
for the string case were 
derived in Ref. \cite{LT}. 
The nonrenormalizable terms can be put in modular invariant form 
by following the procedure outlined in Ref. \cite{KLNI}.
Approximating the Dedekind $\eta$ function by $\eta({\hat T})\approx 
e^{-\pi {\hat T}/12}(1-e^{-2\pi {\hat T}})$, we verified that the
calculation using the modular invariant expression from 
Ref. \cite{LT} (with $\l {\hat T}\r\approx M$)
differ from the results using 
Eq. (\ref{gmcondt}), by at most an order of magnitude. 
The hidden $SU(5)$ matter mass matrix is nonsingular for specific F and D 
flat solutions. In Ref. \cite{SUSYX} a specific cubic level F and D flat 
solution was found. The gravitino mass due to the gaugino and matter 
condensates was estimated to be of the order $1-10~TeV$.
The new aspect of our scenario for supersymmetry breaking is that
supersymmetry 
is broken due to the interplay of the scale generated
by the anomalous $U(1)$ and 
the inclusion of nonrenormalizable terms. Hidden sector
strong dynamics at an intermediate scale may then be responsible for 
generating the hierarchy in the usual way.
Using field theoretic toy models it was demonstrated
that supersymmetry can indeed be broken in this fashion \cite{bd}.
\section{R--parity violation}

Turning to a different issue, recently the H1 and ZEUS collaborations
reported an excess in $e^+P\rightarrow e^+$jet events \cite{h1zeus}.
A possible interpretation 
of this excess is the existence of supersymmetry with
$R$--parity violating couplings. 
This requires that the lepton number violating couplings
are of order one while 
proton lifetime restrictions impose that the baryon number violating
couplings are suppressed
by at least 22 orders of magnitude. Therefore we seek a mechanism
which allows the lepton number violating couplings and forbids the
baryon number violating couplings. 

The model of ref. \cite{custodial} provides an example how
such a mechanism can be realized in superstring theory.
This model gives rise to additional space-time vector
bosons from the basis vectors $\{\alpha,\beta,\gamma\}$ which extend the NAHE
set. These additional vector bosons enhance a combination of the universal and
flavor $U(1)$ symmetries to a custodial $SU(2)$ symmetry. Only the 
Standard Model leptons transform under this $SU(2)$ symmetry
while the quarks are singlets. As a result nonrenormalizable terms of the form
$QLDN$ are allowed by gauge invariance while terms of the form
$UDDN$ and $QQQL$ are forbidden to all orders of nonrenormalizable terms. 
Thus, the VEV of the right handed neutrino in this model induces the desired
lepton number violating coupling while the baryon number violating couplings
are forbidden to all orders of nonrenormalizable terms. 
\section{Beyond superstring}

Next I briefly discuss superstring phenomenology 
in the context of the recently 
discovered superstring dualities.
Superstring theory gives rise to several duality
symmetries \cite{stringdualities}.
The simplest example which illustrates how these dualities arise
is the example of compactification on a circle. When we compactify on a
circle the fact that the wave--function is single valued
imposes that the momenta on the compactified dimension are quantized.
This produces the string momentum modes. 
In string theory in addition we can wrap the string on
the compactified dimension
which produces the winding modes. There is a duality
symmetry which interchanges
momentum and winding modes and at the same time
$R\leftrightarrow1/2R$.
This simple one dimensional example extends to higher dimensions.
In two dimensions
we find three types of duality symmetries. The first which is
geometrical in nature, usually 
referred to as $U$--duality,  the second which exchanges momentum
and winding modes, typically
referred to as $T$--duality and finally a symmetry which
exchanges $T\leftrightarrow U$ and
is known as mirror symmetry. 

The dualities mentioned above are perturbative and exhibit
themselves in the exchange of
the spectrum and the superpotential. Thus they can be
checked order by order in
perturbation theory. In the last few years we have witnessed
a significant progress
in understanding duality symmetries which are nonperturbative,
{\it i.e.}
they exchange weak with the strong coupling.
The starting point in this program
is the Seiberg-Witten solution of $N=2$
supersymmetric pure $SU(2)$ \cite{SW}. 
In the supersymmetric theory the gauge coupling is extended to a complex 
parameter $\tau=\theta/2\pi+i4\pi/g^2$ where $\theta$
is the axial coupling and 
$g$ is the field strength coupling. The strong-weak
duality extends to the full
$SL(2,Z)$ duality of the parameter $\tau$. 
In the Seiberg-Witten solution the exact vacuum structure of the theory
is parameterized in terms of a genus one Riemann surface. 

In string theory we have roughly and very naively a similar situation. The 
gauge coupling is fixed by the VEV of the dilaton field. 
The dilaton field, combined with the space--time components of
the antisymmetric tensor field forms
a modular parameter. In M--theory \cite{witmtheory,Mtheoryreviews}
this complex field is identified with the moduli field of a new dimension 
and hence the $SL(2,Z)$ symmetry of this moduli field translates into
a duality which exchanges strong and weak coupling
\cite{sswd}. In the last couple of
years a large number of qualitative tests have been performed which confirm 
this basic picture. 
The hope that 
understanding nonperturbative string phenomena would reveal 
how a specific vacuum is selected did not materialize so far.
Although the underlying formulation of M--theory
is still unclear, preliminary attempts have been made
to extract some phenomenological information
related to it \cite{mphenomenology}.

The study of nonperturbative dualities indicates a new
structure underlying superstring
theory and quantum space-time.
Underlying superstring and M-theory there should exist
a unifying theory which produces the five known ten
dimensional strings and 11 dimension supergravity
as special limits. 
Currently the popular interpretation is the Matrix
model construction of M-theory \cite{bfss}.
Excellent reviews on these attempts exist in the 
literature \cite{mmreview}.
Here I briefly review my work
with Matone which may (or may not) be related to
this interpretation of quantum 
space-time.

With Matone, inspired by some of the mathematical structure underlying the
recently proposed nonperturbative dualities, we proposed a duality between the
space-coordinate and the
wave function in quantum mechanics \cite{fmone,carroll}.
We suggested that
in quantum mechanics the space coordinate should be interpreted as statistical
variable of a thermodynamical theory underlying the quantum space-time. 
Carrying this interpretation a step further means that the space coordinate
in quantum mechanics should be replaced by a sort of density matrix. 
This interpretation is similar in spirit to the Matrix model
approach to M-theory. 
The expected outcome is that following this interpretation
of the space-coordinate
in quantum mechanics a string-like structure would emerge. 
The reason is that the proposed $\{x-\psi\}$ duality suggests
that second quantization of $\psi$ implies an expansion
of $x$ which is reminiscent of the expansion of $x$ in terms
of the string modes.
However, at present this is still very speculative and not well founded. 

Another view of the recently discovered dualities is that different vacua 
that are classically distinct, are in fact equivalent quantum mechanically. 
Equivalence here means that there exist some coordinate transformation 
that takes one vacuum to another. Recently with Matone we
formulated an equivalence principle in quantum mechanics
\cite{fmtwo}. Starting from the classical
Hamilton-Jacobi equation for the reduced action 
we showed that equivalence under diffeomorphism requires modification 
of the classical equation which is the quantum Hamilton-Jacobi equation. 
This quantum equivalence principle then states that all the quantum
mechanical systems with different potentials are equivalent under a general 
coordinate transformation. The relation of this equivalence principle
to the recently discovered nonperturbative dualities is yet to be uncovered.
\section{Conclusions and outlook}

In summary, string theory provides a window to Planck scale physics
and to the study of the unification of the gauge interactions with gravity.
To bring this exploration from mere speculations into contact with 
experimental physics we have to construct phenomenological superstring models.
The realistic free fermionic models achieved remarkable success in describing 
the real world: existence of three generations with standard $SO(10)$ 
embedding, a superstring solution to the doublet-triplet splitting
problem, correct top quark mass prediction, potentially realistic fermion
mass textures, agreement with $\alpha_s(M_Z)$ and $\sin^2\theta_W(M_Z)$, etc. 
These successes of the realistic free fermionic models may be accidental.
However, taking the view that that is not the case,
the fact that free fermionic
models are formulated at a highly symmetric point in the moduli space
and the underlying $Z_2\times Z_2$ orbifold compactification may be the
origin of their realistic nature. Finally, perturbative string scale
unification 
is possible provided that additional color triplets and electroweak doublets
exist at intermediate energy scales. Such additional states
may give rise to experimentally
accessible signature which may provide 
the smoking gun of superstring unification. 

As an outlook to the future of the phenomenology of superstring
unification it should be 
stressed that the important problem in high energy physics is the nature 
of the electroweak symmetry breaking mechanism. This question must 
be answered by our experimental colleagues and we eagerly
await their resolution of the
issue. With this understanding in mind, and with current experimental
data, supersymmetry and unification are the most
promising theoretical proposals. 
If indeed this is the avenue chosen by nature
then most of the properties
of the Standard Model parameters are determined by some fundamental 
Planck scale theory. Therefore, we must develop the technology needed to study
Planck scale physics. The realistic free fermionic models are designed to
achieve precisely this goal, being the most realistic superstring model
constructed to date. At this time we have only began to scratch the surface 
of this class of models and much further work is required both in order
to develop the calculational tools needed to confront string models with the
experimental data and to better understand
the properties of this class of models.
Understanding the correspondence with other formulations
will provide further insight into their realistic properties,
while experimental findings
of exotic particles may provide further support for their validity. Finally,
it is hoped that these explorations will assist in understanding
if and how a particular string vacuum is selected
and the nature of quantum space-time. 
\section*{Acknowledgments}
I would like to thank the Weizmann Institute and CERN theory group for
hospitality. Work supported in part by DOE grant No. DE--FG--0586ER40272.
\bibliographystyle{unsrt}

\vfill\eject
\end{document}